\documentclass[aps,preprint]{revtex4}  % uncomment, if double spacing is 
\usepackage{graphicx}% Include figure files  
\usepackage{amssymb}

\begin{document}

\preprint{Taraskin et al., Extinction of Epidemics ... }  
  
\title{Extinction of  Epidemics in 
Lattice Models with Quenched Disorder}

\author{S.~N.~Taraskin}  
 \email{snt1000@cam.ac.uk}  
\affiliation{St. Catharine's College and Department of Chemistry, University of Cambridge,  
             Cambridge, UK}  

\author{J.~J.~Ludlam}  
 \email{jjl25@cam.ac.uk}  
\affiliation{Department of~ Plant Sciences, 
University of Cambridge, Cambridge, UK}  
 
\author{C.~J.~ Neugebauer}  
 \email{cjn24@cam.ac.uk}  
\affiliation{Department of Chemistry, University of Cambridge,  
             Cambridge, UK}  

\author{C.~A.~Gilligan}  
 \email{cag1@cam.ac.uk}  
\affiliation{Department of Plant Sciences, University of Cambridge, 
             Cambridge, UK}

\date{\today}% It is always \today, today,  
  
\begin{abstract}  
The extinction of the contact process for epidemics in lattice models with quenched disorder 
is analysed in the limit of small density of infected sites. 
It is shown that the problem in such a regime can be mapped to the 
quantum-mechanical one characterized by the Anderson Hamiltonian for an electron 
in a random lattice. 
It is demonstrated both analytically (self-consistent mean-field) and 
numerically (by direct diagonalization of the Hamiltonian and by means of 
cellular automata simulations) that disorder enhances the contact process 
given the mean values of random parameters are not influenced by disorder. 

\end{abstract}  
  
 
  
\maketitle  
  
%-------------------------------------------------------------------------  
%                       Main text  
%-------------------------------------------------------------------------  
 
%%%%%%%%%%%%%%%%%%%%%%%%%%%%%%%%%%%%%%%%%%%%%%%%%%%%%
\section{Introduction} 
\label{s0}
%%%%%%%%%%%%%%%%%%%%%%%%%%%%%%%%%%%%%%%%%%%%%%%%%%%%% 

The spread of epidemics in complex networks such as biological populations and 
computer networks is of great current interest, both for practical applications
and from a fundamental point of view 
\cite{Hinrichsen_00:review,Odor_04:review,Dorogovtsev_02,Albert_02,Keeling_00:nature,Dybiec_04}. 

This is one of the issues of the theory of non-equilibrium phase transitions 
\cite{Hinrichsen_00:review,Odor_04:review} and the theory of complex networks 
\cite{Dorogovtsev_02,Albert_02}.  
The problems of interest include the question about the existence 
of a critical regime separating invasive (active) and non-invasive (absorbing) 
states of the system and, if such a transition exists, how it depends 
on internal and external parameters and also what the universal features 
of the transition are (see e.g. \cite{Marro_99:book}).  

In one of the simplest models of epidemics, all the nodes are divided into 
two classes: infectious ($I$) and and susceptible ($S$) 
\cite{Liggett_85:book}. 
The epidemic spreads by a contact process according to which 
an infected node can transfer infection to another susceptible 
node with typical 
infection rate $w$ and recover with typical recovery rate $\varepsilon$ 
becoming again susceptible (the SIS model).  
The system undergoes a phase transition with variation of the dimensionless 
parameter, $\eta = w/\varepsilon $, 
from the absorbing  ($\eta < \eta_{\text{c}}$) to  active state 
($\eta>\eta_{\text{c}}$). 
The critical value is  
$\eta_{\text{c}} \sim Z^{-1}$ 
\cite{Marro_99:book,Dorogovtsev_02}, with $Z$ being the typical 
number of links per node (coordination number).  

Usually, the infection and recovery rates are assumed to be node independent. 
However, in real systems, the values of $w$ and $\varepsilon$ can 
vary from node to node (quenched disorder). 
Investigations of contact processes in systems with quenched disorder 
over recent years 
\cite{Hinrichsen_00:review,Odor_04:review,Noest_86,Bramson_91,Moreira_96,Vendruscolo_96,Cafiero_98,Szabo_02,Hooyberghs_03,Hooyberghs_04,Vojta_04} 
have resulted in some rather intriguing findings. 
For example, it has been suggested that the disorder can change 
the universality class of the model \cite{Dickman_98,Janssen_97}. 
However, the situation is far from being completely understood, 
and the aim of this paper is to investigate the influence 
of a general form of quenched disorder on the dynamics of the contact 
process in the absorbing state. 
Using a combination of a simple epidemiological model
with methods from condensed matter physics, we show 
how disorder in the infection or recovery rates, influences  
the long-time dynamics (decay time) of epidemics in the absorbing 
state. This is of practical importance in determining the time to extinction 
of epidemics within this state. 
We also identify a lower bound for $\eta_{\text{c}}$ and show how 
the degree of disorder influences  the magnitude of the extinction rate.

We consider the dynamics of the contact process 
far in the absorbing state when the 
problem can be mapped to the quantum-mechanical one described 
by the disordered Hamiltonian of the Anderson-type 
(see e.g. \cite{Kramer_93}) and an 
approximate method (self-consistent mean-field) can be applied. 
The spectrum of the Hamiltonian under this approximation is then used 
in the analysis of the long-time dynamics of the system. 
The advantage of the approach is in the possibility of
incorporating a general type of disorder in the analysis while 
the disadvantage is due to the rather severe restriction of being 
in the absorbing state (dilute regime for concentration of infected nodes). 
Our main result is that the disorder slows down the long-time dynamics 
of the system given the mean values of the random values stay the same 
as in ordered systems. 
The approximate analytical results are supported by exact numerical 
analysis using a cellular automata approach.  

The paper is organized in the following manner. 
The formulation of the problem is given in Sec.~\ref{s1}. 
The solutions in the dilute regime both for ordered and disordered cases are 
presented in Sec.~\ref{s2} followed by discussion  
in Sec.~\ref{s3}. 
The conclusions are made in Sec.~\ref{s4}. 

%%%%%%%%%%%%%%%%%%%%%%%%%%%%%%%%%%%%%%%%%%%%%%%%%%%%%
\section{Formulation of the problem} 
\label{s1}
%%%%%%%%%%%%%%%%%%%%%%%%%%%%%%%%%%%%%%%%%%%%%%%%%%%%% 

Consider a set of $N$ nodes (sites) connected to each other by 
links (infection paths).  
Each node, $i$, can be in one of two states: infected (occupied 
by an ``excitation'' and 
characterized by occupation number, $n_i=1$) or not infected (empty with 
$n_i=0$). 
The occupation number $n_i$ changes from $0$ to $1$ 
as a result of infection from an occupied node $j$ occurring with  
infection rate $w_{ji}$, and from $n_i=1$ to $n_i=0$ due to natural 
recovery with rate $\varepsilon_i$. 
Any state of the system is characterized by the set of occupation 
numbers, $\{n\}\equiv \{n_1,\ldots,n_N\}$. 
Bearing in mind the stochastic nature of the infection and recovery processes 
it is convenient to characterize the system by the state vector 
$|P(t)\rangle$, the components of which are the probabilities 
of finding the system in different states at time $t$,  
$|P(t)\rangle= |P_{\{n\}}(t)\rangle $, where $n=1,\ldots,2^N$ 
runs over all the possible states of the system. 
The time evolution of the state vector is governed by the 
master equation describing the conserved probability 
flow \cite{Marro_99:book}, 
\begin{equation}
\partial_t |P(t)\rangle = \hat{\mathcal L} |P(t)\rangle~,
\label{e1_1}
\end{equation}
where $\hat{\mathcal L}$ stands for the non-Hermitian 
Liouville operator, the non-zero 
elements of which describe the transitions between the states with 
different numbers of occupied nodes.  

It is convenient to make a 
linear transformation of the state coordinates (change of basis) 
from $P_{n_1,n_2,\ldots,n_N}$ to the $n$-site probabilities, 
\begin{equation}
\overline{P}_{i}(t)=\sum_{n_k\ne n_i}
P_{n_1,\ldots,n_k,\ldots,n_i,\ldots,n_N}(t)~,~~~
\overline{P}_{ij}(t)=\sum_{n_k\ne n_i,n_j}
P_{n_1,\ldots,n_k,\ldots,n_i,\ldots,n_j,\ldots,n_N}(t)~, \ \ \text{etc.}~, 
\label{e1_2}
\end{equation}
where $\overline{P}_{i}(t)$ is the probability of finding node $i$ 
in an occupied ($n_i=1$) state independent of 
the occupation of all other nodes. 
This allows the master equation~(\ref{e1_1}) to be recast  in the following 
form: 
\begin{equation}
\partial_t \overline{P}_{i}(t) = 
-\varepsilon_i \overline{P}_{i}(t) + \sum_{j\ne i} 
w_{ji}(\overline{P}_j(t)- \overline{P}_{ji}(t))  
~, 
\label{e1_3}
\end{equation}
where $\overline{P}_j(t)- \overline{P}_{ji}(t)$ is the probability 
of finding the system 
with the  occupied $j$-th node and the  unoccupied $i$-th node, 
independent of the state of all other nodes. 
The single-site probability $\overline{P}_{i}(t)$ in Eq.~(\ref{e1_3}) 
is coupled with the double-site probabilities  $\overline{P}_{ij}(t)$. 
A similar probability-balance equation for 
$\overline{P}_{ij}(t)$ contains the three-site probabilities 
and so on. 
This makes the set of simultaneous equations to be coupled and thus be  
non-trivial for analysis. 

The lowest level of approximations in decoupling schemes 
involves a complete ignorance of the double-site occupations, 
$\overline{P}_{ij}$, in comparison with other terms in the 
master equation~(\ref{e1_3}), which is possible if 
\begin{equation}
\overline{P}_{ij}(t) \ll \overline{P}_j(t)~\ \ \ 
\text{or} \ \ \ 
\overline{P}_{ij}(t) \ll \frac{\varepsilon_i}{w_{ji}}\overline{P}_i(t)
~, 
\label{e1_4}
\end{equation}
for each pair of communicating sites $i-j$ so that the master 
equation under these approximations transforms to the following form, 
\begin{equation}
\partial_t \overline{P}_{i}(t) = 
-\varepsilon_i \overline{P}_{i}(t) + \sum_{j} 
w_{ji}\overline{P}_{j}(t) 
~,  
\label{e1_5}
\end{equation}
and is hereafter referred to as the approximate master equation. 
The inequalities 
given by Eq.~(\ref{e1_4}) are valid if the typical 
recovery rate is 
much greater than the typical infection rate, $\varepsilon \gg w$,  
and an epidemic dies out 
very quickly over the typical time, $\varepsilon^{-1}$. 
In such a regime, the single-site probabilities are small for 
the majority of sites practically for all times, 
i.e.  $\overline{P}_i \ll 1$, and this regime can be 
called a dilute regime for the concentration of infected sites. 

Therefore, making approximations given by Eq.~(\ref{e1_4}) we focus   
on the dynamics of the system far in the absorbing state
 ( $\eta \ll \eta_c$), i.e. where an epidemic will certainly become extinct. 
Bearing in mind that the terms $\propto \overline{P}_{ij}$ 
(entering  Eq.~(\ref{e1_3}) with a minus sign) 
reduce the infection rate due to a possible 
simultaneous occupation of both communicating sites $i$ and $j$ 
(the transmission of infection cannot occur between two sites if both of them 
are already infected) 
we might expect that the solution of approximate 
rate equation~(\ref{e1_5}) exhibits the enhancement 
of an epidemic in the dilute regime. 
In fact, the approximate master Eq.~(\ref{e1_5}) on its own 
 describes the spread 
of an epidemic in the system of $N$ nodes with multiple reinfection 
of infected nodes where each node can be 
multiply occupied by the ``excitations'' (infection) and 
$\overline{P}_i$ has the meaning of an occupation number which can be larger 
than one. 

The approximate master equation (Eq.~(\ref{e1_5})) similarly to the exact one can also 
have  solutions which behave differently as $t\to\infty$ depending on the 
typical value of the parameter $\eta$. 
In fact, there exists 
a critical value $\eta_{c}^*$ for the approximate master equation which 
separates the absorbing and active states and 
 this critical value  $\eta_{c}^*$ is smaller than the 
critical value $\eta_c$ for the exact master equation due to the nature 
of the approximations made. 
This allows the lower bound estimate, $\eta_{c}^*$,  
for $\eta_c$ to be found by solving 
the approximate problem for a quite general type of disorder.   

The state of the system in the dilute regime can be defined in the subspace 
of the singly-occupied sites spanned by the orthonormal site basis 
 $|i\rangle$, and is characterized 
by the state vector 
$|\overline{P}\rangle = |\overline{P}_1,\overline{P}_2,
\ldots,\overline{P}_N\rangle $ with  components 
$\overline{P}_i(t) = \langle i | \overline{P}(t)\rangle$. 
The master equation~(\ref{e1_5}) can be recast as 
\begin{equation}
\partial_t |\overline{P}(t)\rangle = 
\hat{\mathcal H} |\overline{P}(t)\rangle~,
\label{e1_6}
\end{equation}
where $\hat{\mathcal H}$ stands for the Liouville operator which 
now (under assumption of symmetric infection rates, $w_{ij}=w_{ji}$) 
is Hermitian and can be associated with the 
Anderson-like Hamiltonian (see e.g. \cite{Kramer_93}), 
\begin{equation}
 \hat{\mathcal H} = -\sum_i^N \varepsilon_i |i\rangle\langle i| + 
\sum_{i\ne j}w_{ij} |i\rangle\langle j|~,
\label{e1_7}
\end{equation}
in which  the recovery rates, $\varepsilon_i$, 
play the role of the on-site energies and the 
infection rates, $w_{ij}$, can be associated with the transfer (hopping) 
integrals. 
Both of these values are random and this makes the further analysis 
 non-trivial even in the dilute regime. 
The topology of the underlying network of sites, in principle, 
can be arbitrary but the simplest choice is a regular $D$-dimensional lattice 
with nearest-neighbour interactions only. 
Furthermore, for simplicity, we consider a square lattice and thus the second 
sum in Eq.~(\ref{e1_7}) runs for each site over $Z=4$ its nearest 
neighbours only. 
It is worth mentioning that if the on-site matrix elements 
satisfy the sum rule, $\varepsilon_i = \sum_j w_{ij} $, then 
the number of occupied sites is conserved and the problem 
is equivalent to the random walk problem on a lattice with random 
transition rates \cite{Rudnick_04:book}, or to the scalar 
vibrational problem for a lattice with 
force-constant disorder \cite{Taraskin_02:JPCM}. 
Notice also that decoupling of the single-site probabilities can also be
made in the non-dilute regime by assuming that 
$\overline{P}_{ij}=\overline{P}_{i}\overline{P}_{j}$, i.e. ignoring 
possible correlations in occupation of the communicating sites. 
This brings a non-linearity to the problem which can be treated 
within the  
mean-field approach for ideal lattices \cite{Marro_99:book}.

%%%%%%%%%%%%%%%%%%%%%%%%%%%%%%%%%%%%%%%%%%%%%%%%%%%%%
\section{Solution} 
\label{s2}
%%%%%%%%%%%%%%%%%%%%%%%%%%%%%%%%%%%%%%%%%%%%%%%%%%%%% 

The formal solution of the problem given by Eq.~(\ref{e1_5}) 
is straightforward, 
\begin{equation}
|\overline{P}(t)\rangle = e^{\hat{\mathcal H}t} |\overline{P}(0)\rangle
= \sum_{j}e^{\lambda_j t}\langle {\bf e}^j|\overline{P}(0)\rangle 
|{\bf e}^j \rangle~,
\label{e2_1}
\end{equation}
where $|{\bf e}^j\rangle = |{e}^j_1,\ldots,{e}^j_N\rangle$
 and $\lambda_j$ are the eigenvectors and eigenvalues  
 of the Hamiltonian, respectively, 
$\hat{\mathcal H}|{\bf e}^j\rangle=\lambda_j|{\bf e}^j\rangle $. 
Equivalently, this solution can be written via the Laplace transform 
of the state vector, 
$|\overline{P}(\lambda)\rangle = \int_0^\infty |\overline{P}(t)  \rangle 
e^{-\lambda t}\text{d}t$, 
\begin{equation}
| \overline{P}(\lambda)\rangle = 
\left(\lambda -\hat{\mathcal H}\right)^{-1} |\overline{P}(0)\rangle 
\equiv \hat{\mathcal G} |\overline{P}(0)\rangle
= \sum_{j} (\lambda -  \lambda_j)^{-1} 
\langle {\bf e}^j|\overline{P}(0)\rangle 
|{\bf e}^j \rangle~,
\label{e2_1a}
\end{equation}
where $\hat{\mathcal G}=(\lambda - \hat{\mathcal H})^{-1}$ is 
the resolvent operator. 

We are interested in the time evolution of the total number  of 
infected sites,
\begin{equation}
I(t)= \frac{1}{N} \left\langle \sum_{ii_0}^N 
 \overline{P}_i(t;i_0)\right\rangle \equiv 
 \frac{1}{N} \left\langle \sum_{ii_0}^N 
  \langle i |\overline{P}(t;i_0)\rangle\right\rangle 
~, 
\label{e2_4}
\end{equation}
averaged over different realizations  
of disorder (angular brackets) and/or over initial conditions 
(for concreteness, a single site $i_0$ is infected at $t=0$, 
i.e. $\overline{P}_i(0;i_0)= \delta_{ii_0}$) and 
its Laplace transform, 
\begin{equation}
I(\lambda)=
\frac{1}{N}\sum_{ii_0}^N 
 \langle {\mathcal G}_{ii_0}(\lambda) \rangle
~. 
\label{e2_4a}
\end{equation}
The other quantity of interest is the mean-squared displacement 
of the epidemic, 
\begin{equation}
\langle R^2(t)\rangle= \frac{1}{N} \sum_{ii_0}^N 
{\bf R}_{i_0i}^2 \langle \overline{P}_i(t;i_0)\rangle ~~~
\text{and}~~~
\langle R^2(\lambda)\rangle= \frac{1}{N} \sum_{ii_0}^N 
{\bf R}_{i_0i}^2 \langle {\mathcal G}_{ii_0}(\lambda) \rangle~, 
\label{e2_5}
\end{equation}
where ${\bf R}_{i_0i}$ is the vector connecting site $i_0$ with site $i$. 

As follows from Eq.~(\ref{e2_1}) the dynamics of the system 
in the dilute regime are defined by the eigenspectrum of the 
Hamiltonian. 
The set of characteristic times (inverse eigenvalues) controls the evolution 
of the system with different eigenvalues being important for different 
time scales.  
The long-time dynamics 
of the system are defined by the maximum eigenvalue of the Hamiltonian, 
$\lambda_{\text{max}}$ ($\lambda_{\text{max}}< 0$ in the dilute regime) and our aim 
is to find an estimate for $\lambda_{\text{max}}$ and how it depends on 
the degree of disorder. 
The maximum eigenvalue depends on all the recovery and infection 
rates, $\lambda_{\text{max}}(\varepsilon_i,w_{ij})$, 
and this obviously complicates its analytical evaluation. 
The exact analytical solution of the problem in the general case is 
 not currently known and numerous approximate analytical 
(see e.g. \cite{Kramer_93,Mirlin_00:review}) and numerical 
(see e.g. \cite{Schreiber_96:book,Roemer_03})  
 methods have been developed for evaluation of the spectrum of disordered 
Hamiltonians.  
Below, we use one of the well-developed self-consistent  
mean-field approaches (the 
homomorphic cluster approximation within the coherent potential 
approximation \cite{Yonezawa_79,Li_88}) to find the estimates for 
$\lambda_{\text{max}}$ in the case of off-diagonal disorder 
and compare these results with the exact numerical calculations both for 
the Hamiltonian used in the approximate approach and for the original problem 
(cellular automata (CA) calculations).  
Before analysing the disordered system, we start, however, with a 
trivial case of an ideal crystalline lattice in order to  illuminate our 
approach. 

%%%%%%%%%%%%%%%%%%%%%%%%%%%%%%%%%%%%%%%%%%%%%%%%%%%%%
\subsection{Ideal lattice} 
\label{s2_1}
%%%%%%%%%%%%%%%%%%%%%%%%%%%%%%%%%%%%%%%%%%%%%%%%%%%%% 

In the case of an ideal crystalline lattice, the probability 
distribution functions are $\delta$-functions, 
$\rho_\varepsilon(\varepsilon_i)=\delta(\varepsilon_i-\varepsilon_0)$
and 
$\rho_{w}(w_{ij})=\delta(w_{ij}-w_0) $, and the translationally invariant 
solutions of the eigenproblem are well known 
(see e.g.\cite{Economou_83:book}), so 
that the eigenvectors are the Bloch's waves characterized by the wavevector 
${\bf k}$,    
\begin{equation}
|{\bf e}_{\bf k} \rangle = N^{-1/2} 
\sum_i e^{\text{i}{\bf kR}_i} |i\rangle 
~,
\label{e2_1_2}
\end{equation}
with ${\bf R}_i$ being the position vector of site $i$ 
and the eigenvalues are 
\begin{equation}
\lambda({\bf k}) = -\varepsilon_0 + w_0 S_{\bf k} 
~,
\label{e2_1_3}
\end{equation}
where 
$S_{\bf k}=\sum_j e^{-\text{i}{\bf kR}_{ij}}$ (the sum is taken over $j$ 
running over 
the nearest neighbours to arbitrary site $i$) is the structure factor 
and  
the wavevector ${\bf k}$ lies in the first Brillouin zone 
of the reciprocal space so that  $\lambda_{\text{max}}=\lambda_{{\bf k}=0}= 
-\varepsilon_0+Z w_0$.
  
The Laplace-transform of the total number of infected states 
 is then (see Eq.~(\ref{e2_4a})), 
$
I(\lambda) = 
 (\lambda - \lambda_{{\bf k}=0})^{-1}$, and thus 
\begin{equation}
I(t) = e^{ \lambda_{{\bf k}=0}t}=e^{(-\varepsilon_0 + w_0Z)t}
~.
\label{e2_1_4}
\end{equation}
This means that in the dilute regime, $\eta = w_0/\varepsilon_0 \ll 1 $, 
the exponent in Eq.~(\ref{e2_1_4}) is negative and  the total number 
of infected states decays exponentially with time. 

Therefore, for an ideal crystalline lattice the critical value of 
parameter $\eta=w_0/\varepsilon_0$ obtained 
from equation, $\lambda_{\text{max}}=-\varepsilon_0 +Z w_0=0$, for the system 
described by the approximate master equation
 is 
\begin{equation}
\eta_{\text{cryst}}^* = Z^{-1}
~,
\label{e2_1_4a}
\end{equation}
which gives $\eta_{\text{cryst}}^*=0.25$ for a square lattice with   
nearest-neighbour interactions only.  
This estimate is equivalent to the standard (not self-consistent) 
mean-field estimate, 
 and, as expected, is less than the true critical value, 
$\eta_{\text{cryst}} \simeq 0.4122$ \cite{Marro_99:book}. 

The Laplace transform of the 
mean-squared displacement $\langle R^2(\lambda)\rangle$ for ideal 
crystal in the dilute regime is given by the following expression 
(see Eq.~(\ref{e2_5})), 
$
\langle R^2(\lambda)\rangle  = 
w_0 a^2 Z(\lambda-\lambda_{{\bf k}=0})^{-2}
$, with $a$ being  the nearest neighbour distance, and thus   
\begin{equation}
\langle R^2(t)\rangle =w_0 a^2 Z t e^{ \lambda_{{\bf k}=0}t}= 
w_0 a^2 Z t e^{(-\varepsilon_0 + w_0Z)t}
~.
\label{e2_1_6}
\end{equation}
It follows from Eq.~(\ref{e2_1_6}) that the mean squared displacement 
increases exponentially in the active state when $\eta > \eta_{\text{cryst}}^*$ 
(i.e. $\lambda_{\text{max}}= \lambda_{{\bf k}=0} > 0 $) and 
exponentially decays with time in absorbing state for  $\eta < \eta_{\text{cryst}}^*$ 
(i.e. $\lambda_{\text{max}} < 0 $). 

%%%%%%%%%%%%%%%%%%%%%%%%%%%%%%%%%%%%%%%%%%%%%%%%%%%%%
\subsection{Disordered lattice} 
\label{s2_1a}
%%%%%%%%%%%%%%%%%%%%%%%%%%%%%%%%%%%%%%%%%%%%%%%%%%%%% 

The problems becomes much harder for a disordered lattice characterized 
 by random  infection and recovery rates. 
In order to find the time-dependence of the number of infected sites 
and the mean-squared displacement for the contact process we need 
to evaluate the configurationally averaged resolvent operator, 
$\langle \hat{\mathcal G}\rangle$ (see Eqs.~(\ref{e2_1a})-(\ref{e2_5})). 
This can be done approximately 
 for lattice models with certain types of disorder, 
namely, with diagonal disorder 
(disorder in the recovery rates, $\varepsilon_i$), 
off-diagonal disorder (disorder in transfer rates, $w_{ij}$), and 
for binary systems with substitutional disorder (two species of sites 
randomly occupy the lattice sites \cite{Ehrenreich_76}). 
One of the successful approximate analytical approaches is the self-consistent 
mean-field approach (coherent potential approximation, CPA) 
which allows the main spectral features 
of the disordered Hamiltonian and its eigenfunctions to be modelled 
(see e.g. \cite{Economou_83:book}).  

The main idea of the CPA is in replacement of the disordered lattice 
by the ideal crystalline one which is characterized by the effective 
complex parameters (complex fields), e.g. by the effective recovery, 
$\tilde{\varepsilon}(\lambda)=\tilde{\varepsilon}'(\lambda) + 
\text{i}\tilde{\varepsilon}''(\lambda)$, and transmission, 
$\tilde{w}(\lambda)=\tilde{w}'(\lambda) + 
\text{i}\tilde{w}''(\lambda)$, rates which depend on the eigenvalues, 
$\lambda$, of 
the Hamiltonian and should be found self-consistently. 
The self-consistency equation follows from the requirement that a single 
defect placed in the effective crystal does not scatter the effective 
crystalline eigenfunctions if averaged over disorder. 

In what follows, for concreteness, we consider the case of off-diagonal 
disorder, when all the recovery rates are the same, 
$\rho(\varepsilon_i)=\delta(\varepsilon_i-\varepsilon_0)$,while the 
transfer rates are taken from a uniform (box) distribution, 
\begin{equation}
\rho(w_{ij})= \left\{ 
\begin{array}{cc}
 (2\Delta)^{-1} & ~~~\text{if}~~~ w_0-\Delta \le w_{ij} \le  
w_0+\Delta \\
 0 & ~~~\text{otherwise}
\end{array}
\right.  
~,
 \label{e2_2_1}
\end{equation}
where $\Delta$ is the half-width of the distribution, $0\le \Delta \le w_0$, and the 
mean value  ${\overline w_{ij}}$ coincides with the crystalline one, 
${\overline w_{ij}}=w_0$.  
The particular form of the distribution~(\ref{e2_2_1}) is not important 
for the method discussed below. 
The conclusions are also applicable to any well behaved distribution 
given the mean value coincides with the value for the ordered system.   

The disordered Hamiltonian~(\ref{e1_7}) can be conveniently rewritten 
in the bond representation \cite{Taraskin_02:JPCM}, 
\begin{equation}
 \hat{\mathcal H} = \sum_{(ij)} 
\left(
-Z^{-1}\varepsilon_i |i\rangle\langle i| -  
 Z^{-1}\varepsilon_j |j\rangle\langle j| 
+ w_{ij}|i\rangle\langle j|+ w_{ji}|j\rangle\langle i|\right)
~,
\label{e2_2_2}
\end{equation}
where the summation is taken over all bonds $(ij)$ in the system. 
Such a form of the Hamiltonian allows the single non-correlated 
scatters (bonds) to be introduced in the absence of the on-site disorder 
(the homomorphic cluster approximation \cite{Yonezawa_79,Li_88}). 
The next step is to replace the above Hamiltonian with the effective 
non-Hermitian one, 
\begin{equation}
 \hat{\tilde{\mathcal H}} = \sum_{(ij)} 
\left(
-Z^{-1}\tilde\varepsilon |i\rangle\langle i| -  
 Z^{-1}\tilde\varepsilon |j\rangle\langle j|+  
\tilde{w}|i\rangle\langle j|+\tilde{w}|j\rangle\langle i|\right)
~,
\label{e2_2_2a}
\end{equation}
where the effective fields $\tilde\varepsilon$ and $\tilde{w}$ are 
found from the following two self-consistency equations 
(see Appendix~\ref{app_a}) 
\cite{Chang_87}, 
\begin{equation}
\left\langle
\frac{Z^{-1}(\tilde{\varepsilon}-\varepsilon_0)\pm (w_{ij}-\tilde{w})} 
{1-({\tilde{\mathcal G}}_{ii}\pm {\tilde{\mathcal G}}_{ij})
(Z^{-1}(\tilde{\varepsilon}-\varepsilon_0)\pm (w_{ij}-\tilde{w}))} 
\right\rangle = 0 
~.
\label{e2_2_3}
\end{equation}
The averaging in Eqs.~(\ref{e2_2_3}) is performed over random values 
of transition rates $w_{ij}$  distributed 
according to the probability distribution given by Eq.~(\ref{e2_2_1}). 
The effective resolvent (Green's function) elements 
${\mathcal G}_{ii}$ and  
${\mathcal G}_{ij}$ can be expressed via the ideal crystalline 
resolvent elements, ${\mathcal G}_{ii}^{\text{cryst}}$, of complex argument  
(see Appendix~\ref{app_a}), 
\begin{equation}
{\tilde{\mathcal G}}_{ii}(\lambda) =  
\frac{w_0}{\tilde{w}}{\mathcal G}_{ii}^{\text{cryst}}\left(
\frac{w_0}{\tilde{w}}(\lambda + \tilde{\varepsilon}) - \varepsilon_0
 \right)~, ~~~~~
{\tilde{\mathcal G}}_{ij}(\lambda) = 
\frac{1}{Z\tilde{w}} 
\left[
(\lambda + \tilde{\varepsilon}){\tilde{\mathcal G}}_{ii}(\lambda)-1 
\right]
~,
\label{e2_2_4}
\end{equation}
which are well-known for the square lattice 
(see e.g. \cite{Economou_83:book}). 

%------------------------------------------------------------------------ 
%           Figure 1 
%------------------------------------------------------------------------ 
\begin{figure} % fig 1 
\centerline{\includegraphics[width=13cm]{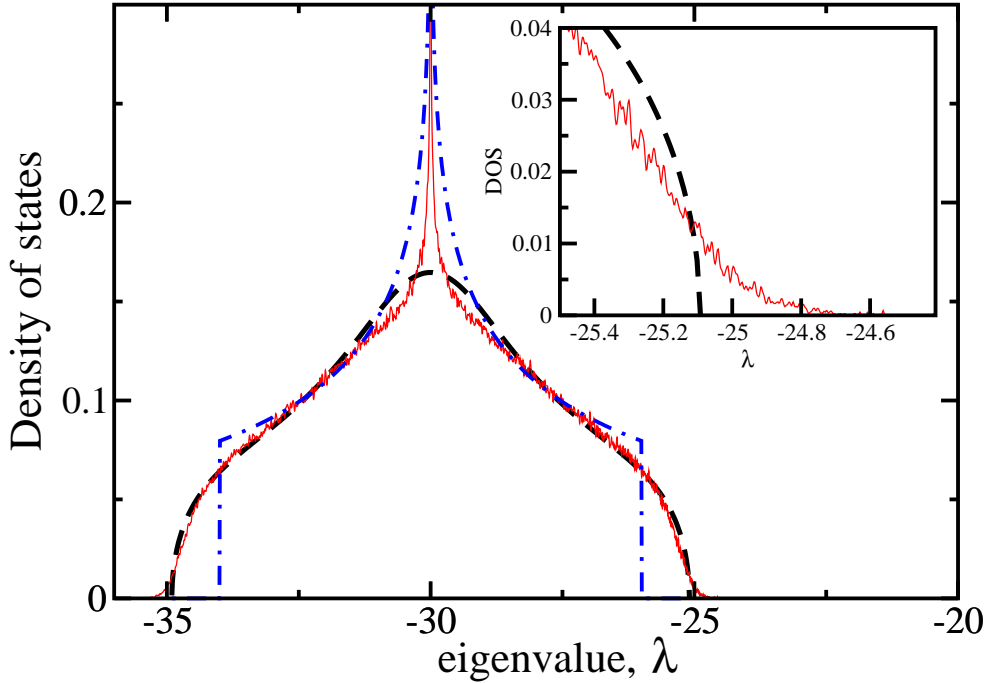}}  
\caption{(Color online) 
The spectrum of the effective (dashed line) and true (solid line) 
 Hamiltonian (density of states) defined on the square lattice 
with nearest-neighbour interactions ($Z=4$) in the 
system with transmission rates uniformly distributed around 
the mean value $w_0 = 1$ with half-width $\Delta=1.0$ and 
$\varepsilon_0=30$. 
The exact spectrum was obtained numerically using the kernel polynomial 
method \cite{Silver_97} for a model of $N=2000\times 2000$ sites. 
The spectrum of the crystalline Hamiltonian with all nearest-neighbour 
interactions ($\Delta=0$) is shown by the dot-dashed line. 
The inset magnifies the spectrum around the top of the band.  
  }  
\label{f1} 
\end{figure}  
%-----------------------------------------------------------------------

The self-consistency equations~(\ref{e2_2_3}) can be solved numerically 
and thus both effective fields can be found. 
Once the complex effective fields, $\tilde{\varepsilon}(\lambda)$ 
and $\tilde{w}(\lambda)$, are known then the spectrum of the 
effective Hamiltonian can be found (see Fig.~\ref{f1}) enabling the dynamics of the system 
in the dilute regime within the self-consistent 
mean-field approach to be studied. 

It can be shown that the total number of infected states and the 
mean-squared displacement in the CPA approximation obey the 
following equations (see Appendix~\ref{app_b}), 
\begin{equation}
I(t) = -\frac{1}{\pi}\text{Im}  
\int_{\lambda_{\text{min}}(\Delta)}^{\lambda_{\text{max}}(\Delta)}
\frac{e^{\lambda t}}{\lambda-\tilde\lambda(\lambda,{\bf k}=0)}
\text{d}\lambda = 
-\frac{1}{\pi}\text{Im}  
\int_{\lambda_{\text{min}}(\Delta)}^{\lambda_{\text{max}}(\Delta)}
\frac{e^{\lambda t}}{\lambda +
\tilde{\varepsilon}(\lambda)-Z \tilde{w}(\lambda)}
\text{d}\lambda
~, 
\label{e2_2_5}
\end{equation}
and
\begin{eqnarray}
\langle R^2(t)\rangle &=& -\frac{1}{\pi}\text{Im}  
\int_{\lambda_{\text{min}}(\Delta)}^{\lambda_{\text{max}}(\Delta)}
e^{\lambda t} 
\left[
\frac{
\nabla_{\bf k}^2 \tilde{\varepsilon}(\lambda,{\bf k})
}
{
(\lambda-\tilde\lambda(\lambda,{\bf k}))^2
}
\right]_{{\bf k}=0}
\text{d}\lambda 
\nonumber 
\\ 
&=& 
 -\frac{1}{\pi}\text{Im}  
\int_{\lambda_{\text{min}}(\Delta)}^{\lambda_{\text{max}}(\Delta)}
\frac{
Za^2 \tilde{w}(\lambda)e^{\lambda t}
}
{
\left( 
\lambda +
\tilde{\varepsilon}(\lambda)-Z \tilde{w}(\lambda)
\right)^2
} 
~\text{d}\lambda  
~,
\label{e2_2_6}
\end{eqnarray}
with the effective dispersion law, 
\begin{equation}
\tilde\lambda(\lambda,{\bf k})= 
-\tilde{\varepsilon}(\lambda)+ \tilde{w}(\lambda)
S_{\bf k}
~.
\label{e2_2_7}
\end{equation}
The integration in Eqs.~(\ref{e2_2_5}) and~(\ref{e2_2_6}) is performed 
over the band(s) of eigenvalues, 
$\lambda_{\text{min}}(\Delta)\le \lambda \le {\lambda_{\text{max}}(\Delta)} $, 
where the imaginary parts of the effective fields are finite for 
$\Delta > 0$. 

As follows from Eqs.~(\ref{e2_2_5}) and~(\ref{e2_2_6}) the long-time 
dynamics of the system both for the number of infected nodes $I(t)$ 
and for the mean-squared displacement $\langle R^2(t) \rangle$ 
are defined by the largest  eigenvalue. 
The upper band edge, $\lambda_{\text{max}}(\Delta)$, can be found within 
the CPA from the self-consistency Eqs.~(\ref{e2_2_3}) by solving them 
for $\lambda > \lambda_{\text{max}}(\Delta)$ where both effective fields are 
real, i.e. $ F_{\pm}(\tilde{w}',\tilde{\varepsilon}',\lambda)=0$ 
with  $F_\pm$ standing for the left hand-side of Eqs.~(\ref{e2_2_3}). 
The analysis of the dependencies of the effective fields on $\lambda$ shows 
that the upper band edge corresponds to the branching point at which 
the following equation holds (see Appendix~\ref{app_c}),   
\begin{equation}
\left[\frac{\partial F_+}{\partial \tilde{w}'}
\frac{\partial F_-}{\partial \tilde{\varepsilon}'} 
- 
\frac{\partial F_-}{\partial \tilde{w}'}
\frac{\partial F_+}{\partial \tilde{\varepsilon}'}
\right]_{\lambda_{\text{max}}(\Delta)}=0 
~.
\label{e2_2_8}
\end{equation}
%
%

%------------------------------------------------------------------------ 
%           Figure 2 
%------------------------------------------------------------------------ 
\begin{figure} % fig 1 
\centerline{\includegraphics[width=13cm]{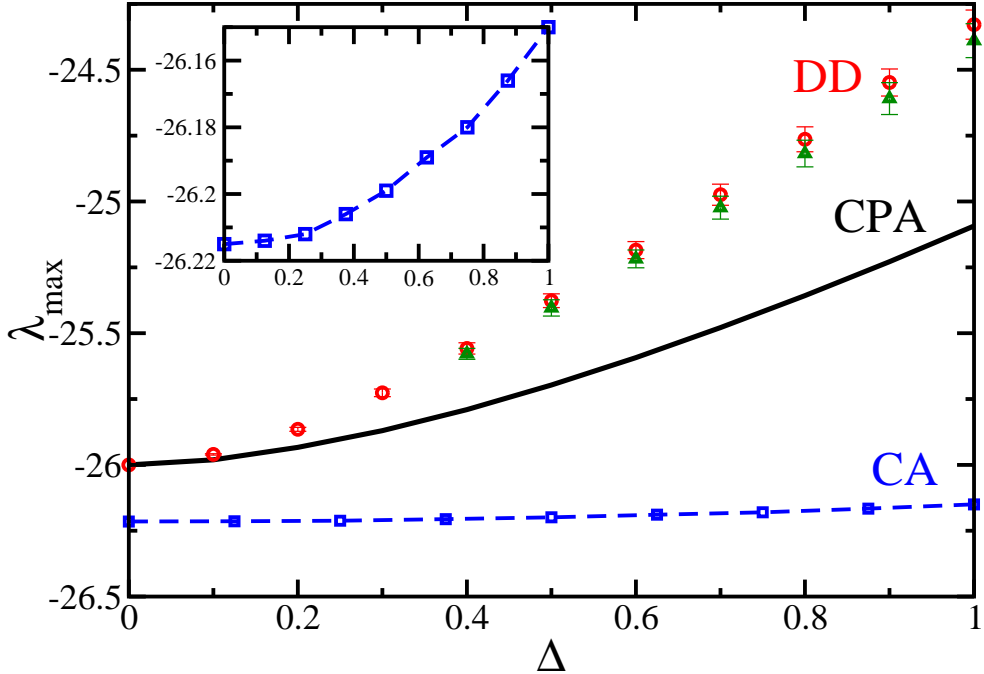}}  
\caption{(Color online) 
The dependence of the maximum eigenvalue, 
$\lambda_{\text{max}}$,  evaluated using a mean-field approach (solid curve,
labelled CPA), 
and  $\lambda_{\text{max}}^*$ calculated by direct diagonalization 
(DD, circles for 4000$\times$4000 and triangles for 2000$\times$2000 lattices; 
the error bars represent the standard deviations of the distribution of 
the maximum eigenvalues),  
 on the degree of disorder 
characterized by the half-width of the box distribution $\Delta$ 
for $\varepsilon_0=30$ and $w_0=1$. 
The squares (labelled CA) represent the long-time decay rates, $\lambda_{\text{CA}}$, 
obtained by the 
CA simulations, each data point corresponding to approximately $5\times 10^{10}$
runs, on a $5\times 5$ lattice. The inset shows a version of the CA data scaled
vertically to clarify the trend.
   }  
\label{f2} 
\end{figure}  
%-----------------------------------------------------------------------

The solution of Eq.~(\ref{e2_2_8}) simultaneously with the self-consistency 
equations~(\ref{e2_2_3}) allows the position of the upper band edge, 
$\lambda_{\text{max}}(\Delta)$, to be found. 
The results of such an analysis are shown in Fig.~\ref{f2} (see the solid line) for 
a particular choice of parameter $\eta \ll \eta_{\text{cryst}}^* \sim 1$.  
When all the transfer rates are the same ($\Delta = 0$), the 
maximum value coincides with the crystalline upper band edge, 
$\lambda_{\text{max}}/w_0=-\eta^{-1}+Z$.
When  the disorder is introduced to the system  
the upper band edge shifts to larger values 
of $\lambda$ and thus the long-time dynamics 
slow down. 
The value of the shift increases with increasing degree of disorder 
characterized by the value of $\Delta$ 
(see Fig.~\ref{f2}). 
This is a general effect that is independent of the type and  form 
(probability distribution functions) of disorder given the mean values 
of random parameters are the same as for the ordered system. 
Indeed, any disorder brought to the system is equivalent to introducing 
additional interactions between the ordered eigenstates which 
unavoidably result in the level-repelling effect for the 
bare (crystalline) eigenstates \cite{Taraskin_01:PRL} 
that is to the broadening of the spectrum. 
Therefore the disorder-induced slowing down of the dynamics of the contact process 
in the dilute regime is a general effect (if disorder does not influence 
the mean values of random variables). 

It is known that the spectrum of the Hamiltonian obtained within the 
self-consistent mean-field approach 
 usually reproduces very well the 
main features of the spectrum of the Hamiltonian (see Fig.~\ref{f1}) 
excluding some special points like singularities (e.g. the mid-band 
singularity) and band edges (see the inset in Fig.~\ref{f1}). 
For the eigenstates around the band edges, the fluctuations 
of random parameters are essential and they lead to 
strong localization of the eigenstates around the band edges. 
In fact, the true (non mean-field) density of states shows
exponentially decaying band tails instead of sharp band edges typical for the 
mean-field crystals (see  the inset in Fig.~\ref{f1}). 
The mean-field approach does not take into account such fluctuations 
and thus the mean-field value of $\lambda_{\text{max}}$ is only the 
(low-bound) estimate of the true maximum eigenvalue of the Hamiltonian, 
$\lambda_{\text{max}}^{*}$. 
The values of $\lambda_{\text{max}}^{*}$ are random values depending 
on a particular realization of disorder. 
We have calculated numerically (using the Lanczos method) the distributions 
of the maximum eigenvalues for system of different sizes 
(up to $N=4000\times 4000$ sites) and  different 
values of $\Delta$. 
The results for $\langle \lambda_{\text{max}}^{*}(\Delta) \rangle$ 
(averaged over $500$ disorder realizations) are shown in Fig.~\ref{f2}. 
The averaged maximum eigenvalue depends on  
$N$ (see Fig.~\ref{f2}) 
apparently approaching some limiting value which is certainly less 
than the band-edge, $-\varepsilon_0 +Zw$, 
for crystal with all transfer rates equal to $w=w_0+\Delta$.

%%%%%%%%%%%%%%%%%%%%%%%%%%%%%%%%%%%%%%%%%%%%%%%%%%%%%
\section{Discussion} 
\label{s3}
%%%%%%%%%%%%%%%%%%%%%%%%%%%%%%%%%%%%%%%%%%%%%%%%%%%%% 

The main finding from our analysis is that the disorder slows down 
the dynamics of the contact process in the dilute regime when the 
system is far in the absorbing state given that the mean values 
of the random parameters are the same as for the ordered system. 
This conclusion is supported by the exact solution for the contact process 
using the cellular automata (CA) simulations 
which were performed using 
a continuous-time algorithm similar to the n-fold way 
(see e.g. \cite{McNeil_94}). 
The CA simulations were used for evaluation of the 
 inverse decay time for $I(t\to\infty)$  
(see squares in Fig.~\ref{f2}) which can be compared with 
$ \lambda_{\text{max}}$ (solid line) and $\lambda_{\text{max}}^{*}$ 
(circles and triangles in Fig.~\ref{f2}). 

First, we compare the approximate (CPA) long-time decay rate 
(magnitude of the maximum eigenvalue)  
 with the exact (CA) one 
for an ideal (ordered) lattice.  
The dilute regime approximation enhances the epidemic 
and thus results in smaller long-time decay rates for $I(t)$ and 
$\langle R^2(t)\rangle$. 
Indeed, as follows from Fig.~\ref{f2} for $\Delta = 0$, the 
exact value $\lambda_{\text{CA}}$ ($\simeq - 26.215$ for a particular 
typical choice of parameters, $\varepsilon_0=30$ and $w_0=1$) is 
smaller than the approximate one 
$\lambda_{\text{max}}=-\varepsilon +Zw_0=-26$. 
When disorder is incorporated in the system the approximate 
decay rate decreases  (eigenvalue increases) 
with increasing disorder (see the solid 
line in Fig.~\ref{f2}). 
The exact value of the decay rate in the disordered system also decreases  
($\lambda_{\text{CA}}$ increases; see the inset in Fig.~\ref{f2})  
with  increasing disorder thus confirming the tendency found in the 
dilute approximation, although the increase in $\lambda_{\text{CA}}$ is 
appreciably smaller than the increase in  $\lambda_{\text{max}}$ and/or 
$\lambda_{\text{max}}^*$.

The value of the decrease in the decay rate with increasing 
disorder, being proportional 
to the width of the tail, 
$|\lambda_{\text{max}}(\Delta=0)-\lambda_{\text{max}}(\Delta=w_0)|
 /w_0 
\sim 
|\lambda_{\text{max}}(\Delta=0)-\lambda^*_{\text{max}}(\Delta=w_0)|/w_0 
\sim 0.1 $, 
is naturally small due to the assumptions made, i.e. 
the system is far in the absorbing state.
However the sign of the effect is important and  
it cannot be predicted by the standard (not self-consistent) mean-field 
analysis for the contact processes \cite{Marro_99:book} if  
the mean values in the disordered system coincide with those for the ordered 
one. 
The other comment concerns the dependence of the effect on the parameters 
of the system. 
The broadening of the spectrum of the Hamiltonian 
does not depend on the mean recovery rate, $\varepsilon_0$. 
This means that if $\varepsilon_0$ increases then the absolute magnitude 
of the effect within the CA treatment, 
$  \lambda_{\text{CA}}(\Delta=0)-\lambda_{\text{CA}}(\Delta=w_0)$ tends 
to  
$\lambda_{\text{max}}(\Delta=0)-\lambda^*_{\text{max}}(\Delta=w_0)$. 
If the recovery rate decreases the system approaches the active state and 
approximations made for the dilute regime break down. 
Therefore the analysis performed cannot be considered as a good approximation
around criticality. 
In fact, as it follows from the preliminary CA analysis, 
the value of 
$  \lambda_{\text{CA}}(\Delta=0)-\lambda_{\text{CA}}(\Delta=w_0)$ 
increases with decreasing $\varepsilon_0$ and can reach zero around 
criticality and even change the sign (to be discussed further elsewhere) i.e. 
the introduction of disorder
into a crystalline system at criticality causes a transition to the absorbing
state rather than further into the active phase.

The last comment concerns a possible rough estimate of the critical 
parameter $\eta_c$ for transition from the absorbing to active 
state in disordered system. 
This estimate can be found by solving the equation 
$\lambda_{\text{max}}(\Delta)=0$ (or $\lambda^*_{\text{max}}(\Delta)=0$), 
which gives, 
$\eta_c \simeq \eta_c^{\text{cryst}} 
(1+(\lambda_{\text{max}}(\Delta=0)-\lambda_{\text{max}}(\Delta=w_0))
 /w_0)$. 
Of course, the quality of this estimate is the same as that for the crystal, 
i.e. $\eta_c^{\text{cryst}}=Z^{-1}=0.25$ as compared to exact value 
$\eta_c\simeq 0.4122$, and can serve only as a reliable low bound for 
the critical value.   

%%%%%%%%%%%%%%%%%%%%%%%%%%%%%%%%%%%%%%%%%%%%%%%%%%%%%
\section{Conclusions} 
\label{s4}
%%%%%%%%%%%%%%%%%%%%%%%%%%%%%%%%%%%%%%%%%%%%%%%%%%%%% 

We have presented the analysis of 
the contact process in the limit of low density 
of occupied (infected) sites (see Eq.~(\ref{e1_4})), i.e. in the 
dilute regime for the infected sites when all the correlation effects 
in occupation probabilities can be ignored.  
This limit occurs, e.g. when the transfer rate is much smaller 
than the infection rate, $w \ll \varepsilon$. 
The system resides in the absorbing state for such a range of parameters 
and its dynamics can be described by using the quantum-mechanical 
tight-binding Hamiltonian. 
The disorder, both in the transfer and recovery rates, can be incorporated 
into the formulation which can be reduced to the eigenproblem for 
the Anderson-like Hamiltonian (see Eq.~(\ref{e1_7})). 
The eigenproblem can be solved approximately analytically 
(self-consistent mean-field) and 
exactly numerically and thus the estimate of the decay rate, 
$|\lambda_{\text{max}}|$,  for long-time 
dynamics  can be found for different degrees of disorder 
(see Fig.~\ref{f2}).
The approximate solution is supported by exact numerical 
analysis using the  cellular automata approach. 
In particular, we conclude that any type of disorder which does not 
change the mean values of random parameters slows down the long-time 
dynamics of the contact process occurring in the far absorbing state. 

\appendix
\section{Self-consistency equation within the homomorphic cluster CPA}
\label{app_a}

In this Appendix, we derive the matrix self-consistency equation within 
the homomorphic cluster CPA. 
Within the self-consistent mean-field approach 
the disordered lattice is replaced by an ideal lattice characterized 
by self-consistently found effective complex parameters (fields), namely, 
by the effective recovery and transmission rates, 
$\tilde{\varepsilon}(\lambda)$ and $\tilde{w}(\lambda)$. 
The effective Hamiltonian describing this effective lattice is given 
by Eq.~(\ref{e2_2_2a}). 
The effective fields are found within the single-defect approximation 
according to the following standard procedure 
\cite{Ehrenreich_76,Economou_83:book}. 
A single defect bond, $(ij)$, taken from the random 
set of bonds characterizing the disordered lattice is placed in the 
effective medium. 
The single-defect Hamiltonian, $ \hat{\tilde{\mathcal H}}_{1}$, is  the 
sum of the ideal effective Hamiltonian and the perturbation due to the 
defect bond, $\hat{\tilde{\mathcal H}}_{1}=\hat{\tilde{\mathcal H}}+
\delta \hat{V}$, where
\begin{equation}
\delta \hat{V}  = -Z^{-1}(\varepsilon_0-\tilde\varepsilon(\lambda))
\left( |i\rangle\langle i|+ |j\rangle\langle j|
\right)
+  
(w_{ij}-\tilde{w}(\lambda))
\left(|i\rangle\langle j|+|j\rangle\langle i|\right)
~,
\label{app_a_1}
\end{equation}
This defect bond influences (scatters) the eigenfuctions of the 
original effective Hamiltonian. 
In the CPA, the effective medium is tuned in such a manner that 
this scattering vanishes on average. 
In other words, the single-defect scattering operator $\hat{T}$, 
introduced by equation 
$\hat{T}=\delta\hat{V} + \delta\hat{V}\hat{\tilde{\mathcal G}}\hat{T}$ 
(where $\hat{\tilde{\mathcal G}}=(\lambda - \hat{\tilde{\mathcal H}})^{-1}$ is the 
effective resolvent) averaged over different realizations of defect bond 
taken from the same probability distribution as for disordered medium 
should be zero, i.e. 
\begin{equation}
\langle \hat{T}\rangle = 
\langle \delta \hat{V}(1-\hat{\tilde{\mathcal G}}\delta \hat{V})^{-1}
 \rangle =0
~,
\label{app_a_2}
\end{equation}
The above equation is the self-consistency matrix equation where 
the scattering matrix in the site basis is 
\begin{equation}
 \hat{T} = \frac{1}{|1-{\tilde{\mathcal G}}\delta \hat{V}|} 
\left(
\begin{array}{c c}
\delta\varepsilon + (\delta w^2-\delta\varepsilon^2)
{\tilde{\mathcal G}}_{jj}
 &~~~ 
\delta w - (\delta w^2-\delta\varepsilon^2)
{\tilde{\mathcal G}}_{ij} 
\\ 
\delta w - (\delta w^2-\delta\varepsilon^2)
{\tilde{\mathcal G}}_{ji} 
&~~~ 
\delta\varepsilon + (\delta w^2-\delta\varepsilon^2)
{\tilde{\mathcal G}}_{jj}
\end{array}
\right)~, 
\label{app_a_3}
\end{equation}
with $\delta\varepsilon= -Z^{-1}(\varepsilon_0-\tilde\varepsilon(\lambda))$ 
and $\delta w= w_{ij}-\tilde{w}(\lambda)$. 
Bearing in mind that 
${\tilde{\mathcal G}}_{ii}={\tilde{\mathcal G}}_{jj}$ and
${\tilde{\mathcal G}}_{ij}={\tilde{\mathcal G}}_{ji}$ the scattering 
matrix can be easily diagonalized by a similarity transformation 
to a new basis, so that 
\begin{equation}
 \hat{T} = 
\left(
\begin{array}{c c}
\frac{\delta\varepsilon - \delta w}
{1-({\tilde{\mathcal G}}_{ii}-{\tilde{\mathcal G}}_{ij}) 
(\delta\varepsilon - \delta w)}
 & 
0 
\\ 
0
& 
\frac{\delta\varepsilon + \delta w}
{1-({\tilde{\mathcal G}}_{ii}+{\tilde{\mathcal G}}_{ij}) 
(\delta\varepsilon + \delta w)}
\end{array}
\right)~, 
\label{app_a_4}
\end{equation}
which straightforwardly results in Eq.~(\ref{e2_2_3}). 

The elements of the scattering matrix $\hat{T}$ depend on the diagonal 
and off-diagonal matrix elements of the effective resolvent, 
${\tilde{\mathcal G}}_{ii} $ and ${\tilde{\mathcal G}}_{ij}$, 
respectively. 
The effective Hamiltonian describes the ideal lattice characterized 
by complex parameters, $\tilde\varepsilon$ and $\tilde w$, and thus its 
eigenfuctions are the Bloch's waves given by  Eq.~(\ref{e2_1_2}) with 
effective dispersion described by  Eq.~(\ref{e2_2_7}). 
This allows the real-space matrix elements of
the resolvent to be expressed via the reciprocal-space ones, and then 
via similar elements of the 
resolvent for ideal crystalline lattice,   
\begin{eqnarray}
{\tilde{\mathcal G}}_{ii}(\lambda)&=& 
\frac{1}{N}\sum_{\bf k}
\frac{1}
{\lambda +\tilde\varepsilon -\tilde w S_{\bf k}} =
 \frac{w}{\tilde{w} N}\sum_{\bf k}
\frac{1}
{\lambda w/\tilde{w} +\tilde\varepsilon w/\tilde{w}  
-\varepsilon_0 - \lambda_{\text{cryst}}(\bf k)}
\nonumber
\\
&=& \frac{w}{\tilde{w}}{\mathcal G}_{ii}^{\text{cryst}}
\left[\frac{w(\lambda+\tilde\varepsilon)}{\tilde{w}} - \varepsilon_0
\right]
\label{app_a_5}
\end{eqnarray}
with ${\mathcal G}_{ii}^{\text{cryst}}(\lambda)= 
N^{-1}\sum_{\bf k}(\lambda -\lambda_{\text{cryst}}({\bf k}))^{-1}
$ being the crystalline resolvent characterized by 
 the crystalline dispersion, 
$ \lambda_{\text{cryst}}(\bf k)$,  
given by Eq.~(\ref{e2_1_3}) 
and 
\begin{eqnarray}
{\tilde{\mathcal G}}_{ij}(\lambda)
&=& 
\frac{1}{N}\sum_{\bf k}
\frac{e^{-\text{i}{\bf k}{\bf R}_{ij}}}
{\lambda -\tilde\lambda(\lambda,{\bf k})} =
 \frac{1}{\tilde{w} Z N}\sum_{\bf k}
\left(
\frac{\lambda +\tilde\varepsilon}
{\lambda -\tilde\lambda(\lambda,{\bf k})}-1
\right)
\nonumber
\\
&=& 
\frac{\lambda +\tilde\varepsilon}{\tilde{w} Z}
\frac{1}{N}\sum_{\bf k}\frac{1}{\lambda -\tilde\lambda(\lambda,{\bf k})}
-\frac{1}{\tilde{w} Z}~, 
\label{app_a_6}
\end{eqnarray}
where $N^{-1}\sum_{\bf k}(\lambda -\tilde\lambda(\lambda,{\bf k}))^{-1}=
{\tilde{\mathcal G}}_{ii}(\lambda)$. 
Eqs.~(\ref{app_a_5})-(\ref{app_a_6}) justify the expressions for the 
matrix elements of the effective resolvent given by Eq.~(\ref{e2_2_4}).

\section{Number of infected sites within the CPA}
\label{app_b}

The aim of this appendix is to derive Eq.~(\ref{e2_2_5}) for the number 
of infected sites as a function of time. 
It is convenient to find the Laplace transform, $I(\lambda)$, 
of the function $I(t)$ and 
then use the inverse transform, 
\begin{equation}
I(t)=\frac{1}{2\pi\text{i}}
\int_{\delta-\text{i}\infty}^{\delta+\text{i}\infty}e^{\lambda t}I(\lambda)
\text{d}\lambda
~, 
\label{app_b_1}
\end{equation}
to reveal Eq.~(\ref{e2_2_5}). 
The Laplace transform $I(\lambda)$ is given by Eq.~(\ref{e2_4a}) 
in which, within the CPA, 
the averaged resolvent matrix element should be replaced 
by the matrix element of the effective resolvent, 
\begin{equation}
I(\lambda)=
\frac{1}{N}\sum_{ii_0}^N 
 \tilde{\mathcal G}_{ii_0}(\lambda)=
\frac{1}{N}\sum_{ii_0}^N 
\frac{1}{N}\sum_{\bf k}
\frac{e^{-\text{i}{\bf k}{\bf R}_{ii_0}}}
{\lambda -\tilde\lambda(\lambda,{\bf k})}
~. 
\label{app_b_2}
\end{equation}
Bearing in mind the identity $\sum_i e^{-\text{i}{\bf k}{\bf R}_{ii_0}}=
N\delta_{{\bf k},0}$, we obtain 
$ I(\lambda)=
(\lambda -\tilde\lambda_{{\bf k}=0})^{-1}$. 
Substitution of this expression in Eq.~(\ref{app_b_1}) gives
\begin{equation}
I(t)=\frac{1}{2\pi\text{i}}
\int_{\delta-\text{i}\infty}^{\delta+\text{i}\infty}\frac{e^{\lambda t}}
{\lambda -\tilde\lambda_{{\bf k}=0}}
\text{d}\lambda
~.  
\label{app_b_3}
\end{equation}
The effective dispersion, $\tilde\lambda_{\bf{k}}$, depends on the 
 the effective fields $\tilde{w}(\lambda)$ 
and $\tilde\varepsilon(\lambda)$ which are analytic functions of $\lambda$ 
everywhere 
except the finite interval on the real axis (branch cut), 
$\lambda \in [\lambda_{\text{min}},\lambda_{\text{max}}]$, where the 
density of states of the effective Hamiltonian is finite 
(see e.g. \cite{Gonis_77} and references therein). 
The contour of integration in Eq.~(\ref{app_b_3}) can be transformed into 
a closed one around the branch cut and thus,  
taking into account that the real part of the integrand is a continuous function
through the branch cut but the imaginary part changes sign, 
Eq.~(\ref{e2_2_5}) follows from Eq.~(\ref{app_b_3}) and Eq.~(\ref{e2_2_6}) 
can be derived in a similar fashion. 

\section{Equation for the band edge}
\label{app_c}

The two self-consistency Eqs.~(\ref{e2_2_3}) can be recast 
in the following form: 
\begin{eqnarray}
\int_{0}^{\infty}
\frac{\delta\varepsilon - \delta w}
{1-({\tilde{\mathcal G}}_{ii}-{\tilde{\mathcal G}}_{ij}) 
(\delta\varepsilon - \delta w)}\rho(w_{ij})\text{d}w_{ij} 
&\equiv& F_{-}(\tilde\varepsilon,\tilde{w},\lambda)=0
\label{app_c_1} 
\\
\int_{0}^{\infty}
\frac{\delta\varepsilon + \delta w}
{1-({\tilde{\mathcal G}}_{ii}+{\tilde{\mathcal G}}_{ij}) 
(\delta\varepsilon + \delta w)}\rho(w_{ij})\text{d}w_{ij} 
&\equiv& F_{+}(\tilde\varepsilon,\tilde{w},\lambda)=0
~. 
\label{app_c_2}
\end{eqnarray}
For any value of $\lambda$ in the complex plane, these equations can be 
solved and two complex fields, $\tilde\varepsilon(\lambda)$ 
and $\tilde{w}(\lambda)$, can be found. 
On the real axis for $\lambda$, $\lambda\ge \lambda_{\text{max}}$,  
the both fields are also real, $\tilde\varepsilon(\lambda)=
\tilde\varepsilon'(\lambda)$ 
and $\tilde{w}(\lambda)=\tilde{w}'(\lambda)$ with $\lambda_{\text{max}}$ 
being the branching point. 
In this range of $\lambda$, it is convenient to rewrite 
Eqs.~(\ref{app_c_1})-(\ref{app_c_2}) in the following form:
\begin{equation}
\left\{
\begin{array}{c c}
\tilde\varepsilon'&=\varepsilon_{-}(\tilde{w}',\lambda)
\\ 
\tilde\varepsilon'&=\varepsilon_{+}(\tilde{w}',\lambda)
\end{array}
\right.~, 
\label{app_c_3}
\end{equation}
where $\varepsilon_{\mp}$ are multivalued functions of $\tilde{w}'$ for fixed 
$\lambda$. If $\lambda > \lambda_{\text{max}}$ these contour lines usually 
cross at two points one of which corresponds to the physical solution. 
At the branching point, $\lambda = \lambda_{\text{max}}$, these two 
solutions merge and the condition for this is
\begin{equation}
\frac{\partial\varepsilon_{-}}{\partial \tilde{w}'} = 
\frac{\partial\varepsilon_{+}}{\partial \tilde{w}'}
~, 
\label{app_c_4}
\end{equation}
which together with Eqs.~(\ref{app_c_1})-(\ref{app_c_2}) or with 
 Eqs.~(\ref{app_c_3}) allows the location of the upper band, 
$\lambda_{\text{max}}$, to be found. 
Eq.~(\ref{app_c_4}) can be rewritten in the more elegant but equivalent form 
given by Eq.~(\ref{e2_2_8}). 
Indeed, differentiation of Eqs.~(\ref{app_c_1})-(\ref{app_c_2}) with respect 
to $\tilde{w}'$ gives 
\begin{equation}
\frac{\partial\varepsilon_{\mp}}{\partial \tilde{w}'} = 
-\frac{\partial F_{\mp}}{\partial \tilde{w}'}
\left(
\frac{\partial F_{\mp}}{\partial \tilde{\varepsilon}'}
\right)^{-1}
~, 
\label{app_c_5}
\end{equation}
from which Eq.~(\ref{e2_2_8}) follows straightforwardly with the use 
of Eq.~(\ref{app_c_4}).

%\bibliography{archive_snt} 

\end{document}